\documentstyle[A4,11pt,psfig]{article}
\begin{document}
\title{A consistent interpretation of the KARMEN anomaly}
\author{J.-M. L\'evy and F. Vannucci \thanks{Laboratoire de Physique Nucl\'eaire et de Hautes Energies,
CNRS - IN2P3 - Universit\'es Paris VI et Paris VII, Paris.   \it Email: jmlevy@in2p3.fr , vannucci@in2p3.fr}}
\pagenumbering{arabic}
\sloppy
\maketitle
\begin{abstract}
The Karmen anomaly can be interpreted as being due to a heavy neutrino
of mass around 137 MeV and mean life 14 s produced in $\pi_{e 2}$ decays. This interpretation is 
consistent with the present limits on the couplings of such an object.
\end{abstract}
\newpage
\begin{center}
%\LARGE{\underline{Introduction.}}
\end{center}
\par
\section{\Large What is the KARMEN anomaly ?}

The Karmen collaboration has reported a distortion in the time distribution of
their events  \cite{Kar}. The experiment studies low energy neutrinos produced
by the dump of $800$ MeV protons at the ISIS facility of the Rutherford
Laboratory.  The accelarator has a unique time structure which allows to
discriminate between neutrinos coming from $\pi$ decaying at rest and from the
subsequent $\mu$ decay at rest. It is in this last process that an anomaly was
observed in the time distribution of interactions, which overall reproduces the
lifetime of the muon but for an excess of events at a time delay centered on
$3.6 \mu$s w.r.t. the arrival of the protons. The signature is a
deposition of energy in the range 11 \rm to 35 MeV.\\
 This is interpreted by the Karmen collaborators as possibly coming from the
production of a heavy neutral object $X$ of mass 33.9 MeV, travelling with a
$\beta = v/c$ of $1/60$ and decaying in the detector, $17.5 m$ away from the
dump. The $X$ particle is produced in $\pi$ decays : $\pi \rightarrow \mu {X}$
with a very small branching ratio and  decays in turn into the {\it a priori}
open channels: $X \rightarrow \nu \gamma$ or $X \rightarrow \nu \rm{e}^+
\rm{e}^-$ or $X \rightarrow  \nu \gamma \gamma$\\

The Karmen experiment finds about 100 events compatible with $X$ production and decay in a
sample of 3000 neutrino interactions. This is a more than 4 $\sigma$ effect. \\
Experimentaly, there are two unknowns in the problem: the branching fraction $Br$ of $\pi$'s
decaying into the $X$ particle and the lifetime $\tau$ of the latter. In order
to reproduce the observed number of events, $Br$ and $\tau$ must be related by
the curve shown in Fig.1 as given in \cite{Kar}.
For long lifetimes, the
probability of decay in the detector is inversely proportionnal to the lifetime.
For shorter lifetimes, a good fraction of $X$'s has decayed before reaching the detector.
Observe that, for very short $\tau$, the distribution
of excess events in the detector should be uneven because of the stronger probability to decay
close to the upper end. Taking this into account would change the shape of the curve in the low
$\tau$ region. However, Karmen sees no such effect, which can be taken as an indication that of the
two possible $\tau$ values corresponding to a given $Br$, only the higher one must be considered.\\
A direct search for $X$ production in $\pi_{\mu 2}$ decay by looking at the
muon spectrum was  performed at PSI \cite{PSI}, \cite{Dau}. It excludes
branching fractions larger than $2.6 \times 10^{-8}$, as also
indicated in Fig. 1.\\

\section{\Large Production and decay of massive neutrinos}
Massive neutrinos, arising, for example, from the introduction of right-handed
chiral states, occur naturally  in most extensions of the standard model. They
have been searched through their decays in various experiments \cite{Greg,JM}\\
\subsection{Production}
If massive neutrinos exist, the so-called weak eigenstates (associated to
charged leptons via weak currents) which we will call generically
$\nu_{\lambda}$, are linear combinations of mass eigenstates $\nu_h$ . Let us
call $U_{h \lambda}$ the coefficient of $\nu_h$ in the expansion of
$\nu_{\lambda}$ ($\lambda$ being  $e$ or $\mu$ for the production processes
envisionned here). In any '$\nu_{\lambda}$' neutrino beam, there is a $\nu_h$
component at the $|U_{h \lambda}|^2$ level up to phase space factors and weak
currents matrix elements.\\ For example, $\pi_{e 2}$ decays may yield $\nu_h$
at the level $|U_{h e}|^2$ up to the above mentionned factors, provided that
$\nu_h$ has a mass below $m(\pi)-m(e) = 139 $ MeV, while $\pi_{\mu 2}$ decays
would produce $\nu_h$ at the level $|U_{h \mu}|^2$ with masses up to
$m(\pi)-m(\mu) = 34 $ MeV. \\Note that the well known chirality suppression of
$\pi_{e 2}$ no longer occurs as soon as $m_{\nu} \approx $ a few MeV .\\ For
the $X$ particle, $\lambda = \mu$ and the mixing discussed in Fig 1 is $U_{h
\mu}$ . Since $m_{X}$ is supposed to be at the limit of the available
phase-space in $\pi_{\mu 2}$ (in order to ensure $\beta = 1/60$), one would
expect it to be produced more abundantly in $K_{\mu 2}$. The ratio of the branching
ratios $\frac{Br(K \rightarrow \mu X)}{Br(\pi \rightarrow \mu X)} \approx 18$
instead of $.63$ for massless $\nu$'s. However, this is very sensitive to the precise
value of $m_X$.

\subsection{Decays}
Several decay modes can be investigated. Heavy neutrinos with masses below $1 $
MeV can only decay radiatively by emitting one or two photons. As soon as the
mass is above $1 $ MeV, the most favoured decay mode is $e^+ e^- \nu_{e}$. The
matrix element is analogous to that of muon decay and the width is given by
$$\Gamma(\nu_h \rightarrow e^- e^+
\nu)=\frac{G_F^2}{192\pi^3}m_{\nu_h}^5f(m_{\nu_h}) |U_{h e}|^2$$ 
where $f$ tends rapidly to $1$ for masses above $10 $ MeV (\cite{JM}) and $|U_{h e}|$
is the same mixing which appears at production if $\nu_h$ is produced together
with an electron.\\ For higher masses, other decay channels open: $\nu e^{\pm}
\mu^{\mp}, e^- \pi^+, \mu^- \pi^+, \nu \mu^- \mu^+ ...$ each one involving a
mixing matrix element which depends on the final state leptons.

\section{\Large $X$ or $X^*$ particle?}
The $X$ particle could be such a massive neutrino undergoing weak decays.
However for a $33.9 $ MeV object $e^+ e^- \nu_e$ is the only open non-radiative
channel and its width, far above those of the radiative
modes, dominates the lifetime.
An upper limit of $|U_{h e}|^2 \leq 5 \times 10^{-6}$ has been obtained
\cite{Greg} for such a mass. This translates into a limit $\tau \geq 120 \ s$
which is reported in Fig. 1 and excludes most of the allowed parameter region.
For the highest mass eigenstate,  published limits on $\Gamma(\pi \rightarrow e
\nu)/\Gamma(\pi \rightarrow \mu \nu)$ (\cite{Brit})
allow to infer $|U_{h e}|^2 \leq 1.23 \times 10^{-6}$ and $\tau \geq 500 \ s$\\
Moreover, taking the experimental Karmen curve (see Fig. 1) into account, this implies \\
$Br(\pi^+ \rightarrow \mu^+ X) \geq 4.5 \times 10^{-9}$ which is barely
compatible with the upper limits  from the two PSI groups who tried a direct
$X$ search \cite{PSI}, \cite{Dau}) \footnote{As quoted in the Review of
Particle Physics \cite{PDG} the PSI results would seem to contradict the
lower bound we have just given and rule out altogether the Karmen solution.
However the PDG are wrong in that they call limits on $|U_{X \mu}|^2$ figures
which are really limits on the branching ratio. Since, on the other hand, the
way the PSI groups compute their upper bounds does not yield the tightest
possible result, their is ample room for argument, and very little room,
if any, for $X$} \\

An alternative interpretation has been proposed \cite{Gnin} in terms of exotic muon decays. It
involves unknown scalars. We study here yet another possibility, which implies an 'ordinary'
massive neutrino, call it $X^*$.\\

As already remarked, many channels open as soon as phase-space permits. If we
assume a $\pi_{e 2}$ production mode, masses up to $139 $ MeV are possible, and
$\nu e^{\pm} \mu^{\mp}$ decays are allowed for sufficiently high $m_{X^*}$, as
schematically shown in Fig. 2 . For such modes, the time delay between beam on
target and the events in excess in the Karmen detector would be the sum of the
final state muon lifetime with the time necessary for $X^*$ to fly from the
proton target to the detector.  Indeed, the kinetic energy of $\mu^{\pm}$
is small and that of $e^{\mp}$ may fall short of the Karmen detection energy
threshold, depending on $m_{X^*}$, rendering it necessary to wait for the
muon to decay. \footnote{for $X^*$ at rest, one finds $\overline{E}_e \leq 14$
MeV and $\overline{T}_{\mu}\leq 2$ MeV.}\\
This would yield an average (electron) energy around $36$ MeV after an average delay of $2.2 \mu s$
following the undetected $X^*$ decay.\\
The Karmen anomalous events appear to be rather tightly clustered in time if they are interpreted
in the lines of \cite{Kar} and their apparent dispersion cannot be understood as due to a
muon decay. However, the authors of \cite{Gnin} show that a rather narrow bump
can be interpreted as the superposition of two exponentials, both with the same (muon) decay
time constant, but with the second shifted by a more or less fixed amount representing a time
of flight.\\
\subsection{Model for the time distribution}
Therefore, we have tried to perform a two component fit to the Karmen time distribution.
The first component is an exponential reflecting ordinary muon decay in the target followed
by practically instantaneous interaction of the normal light neutrinos thus produced in the detector
(the ToF is about 60 ns)\\ The second component is the convolution
of a uniform and an exponential distribution representing the sum of the transit time of a heavy neutrino
from the target to the detector (which is also the heavy neutrino lifetime in the lab and is practically
uniform in the window of interest) plus the lifetime of the muon produced in the heavy neutrino decay. \\
However, this turned out to be inconsistent. The fitted ToF of the heavy neutrinos implies a mass close
to $137$ MeV which in turn means a positron energy distribution (in $X^* \rightarrow \mu^- e^+ \nu$) rather
evenly distributed around an expectation value of $14$ MeV, thereby invalidating the hypothesis that the delay
would always comprise the $\mu$ lifetime.
It was therefore necessary to add a third component representing those 'direct' positrons.
Thus, there are altogether 5 parameters
representing the intensities of the three components and the lower and upper limits of the time window
corresponding to the passage of the $X^*$'s in the Karmen detector.\\
In principle, this gives us a potential consistency check of the fact that the anomaly is really due to a
particle since its speed can be estimated from the last two parameters using the average distance between the
proton target and the detector on the one hand, and the length of the latter on the other hand.
However, the data at our disposal are not accurate enough to decide between different fit results which
do or do not conform with this expectation. Given the large errors on the fitted parameters, the results
presented on Fig. 3 cannot be considered as inconsistent with  the expectation.\\
Here, $T1$ and $T2$ are the lower and upper edges of the time window, and the three following items are 
constants multiplying the three normalized distributions superposed in the fit. Since 
the bin width is $.5 \mu s$, the corresponding number of $\mu^- \rightarrow e^-$ and $X^* \rightarrow e^+$
are about $66$ and $38$ respectively.\\
The fit yields: ToF $=3.66 \mu s$, hence $\beta = .0159$ and $m_{X^*} 
\approx 137.3 $ MeV.\\
\subsection{Interpretation}
It must be remarked now that for an 'ordinary' heavy Dirac neutrino, there are really two (non mutually exclusive)
possibilities:\\
- either it decays to $e^-$ and a virtual $W^+$ which yields $\mu^+ \nu_{\mu}$\\
- or it decays to $\mu^-$ and the virtual boson goes into $e^+ \nu_e$.\\
The first possibility involves the same mixing matrix element, namely $U_{h e}$, which appears at
production, whilst in the second case, the decay is due to $U_{h \mu}$\\
Clearly enough, in any case $U_{h e} \neq 0$ and the $e^+ e^- \nu_e$ mode is also widely open.
If we assume the first channel, the unknown mixing cancels in the ratio of the two modes which is
given by the following formula:$$\frac{\Gamma(e^- \mu^+ \nu_{\mu})}{\Gamma(e^- e^+ \nu_e)} =
(1-8r+r^2)(1-r^2)-12rLog(r)$$ where $r= (m_{\mu}/m_{X^*})^2 $ and $m_e=m_{\nu_e}=m_{\nu_{\mu}}=0$ \cite{JM}.
The width ratio is about $0.5\%$ for $m_{X^*} = 137$ MeV.\\
This shows two things: first, for the range of masses of interest, the lifetime is dominated
by the $e^+ e^- \nu_e$ mode, which gives us another relation between $|U_{h e}|$ and $\tau$ and allows
to draw further conclusions. Second, decays through this mode  must have occured abundantly in the
Karmen experiment, but
the corresponding (large) deposition of energy is probably rejected by the Karmen selection.\\
The results of this analysis are summarized on Fig. 4 It reproduces the Karmen 'efficiency' curve
corrected for the $\pi$ decay mode considered here (The unknown constant was taken from the
asymptotic direction of the Karmen plot \cite{Kar}); also drawn is the branch of hyperbola corresponding
to the $Br \times \tau $ value fixed by the $X^*$ mass and the hypothesis of an $e^- e^+ \nu$
dominated $X^*$ lifetime. The latter is found to be $\tau_{X^*} \approx 14 s$ and the
$\pi^+ \rightarrow e^+ {X^*}$ branching ratio  is around $4 \times 10^{-10}$ This entails a mixing matrix element
$|U_{{X^*} e}|^2 \approx 4.2 \times 10^{-8}$ compatible with the previously published upper limits \cite{Greg,JM}\\
Here again,this object should be much more abundantly produced in kaon decays. The 'ratio of ratios' is
around $100$ in this case.\\
Of course, we do not claim that this is "the" solution, but only wish to point out another possibility. 
It could probably be easily disproved through confrontation with more detailed Karmen data.
\section{\Large Conclusion}
In short, the Karmen collaboration have chosen to explain their anomalous events by assuming they originate
from a rare $\pi_{\mu 2}$ decay, because a 'neutrino' with a mass at the edge of phase-space (around
$34 $ MeV) has both a low enough velocity to explain the $3.6 \mu s$ delay and a
low enough mass to explain the $30 $ MeV energy deposition. We claim 
that both features can be explained by a $\pi_{e 2}$ decay yielding a much
heavier 'neutrino' which decays itself through a $e^{\pm} \mu^{\mp} \nu_l$ channel,
the signature of which is a visible energy in the range of
interest after the (apparently) observed delay.
The more mundane $e^{\pm} e^{\mp} \nu_l$ channel is supposedly hidden by the experimental
energy cut off and the apparent clustering in time due to a fixed time of flight followed by a
short detectability window is not spoiled upon taking the muon lifetime into account.
\newpage
\psfig{file=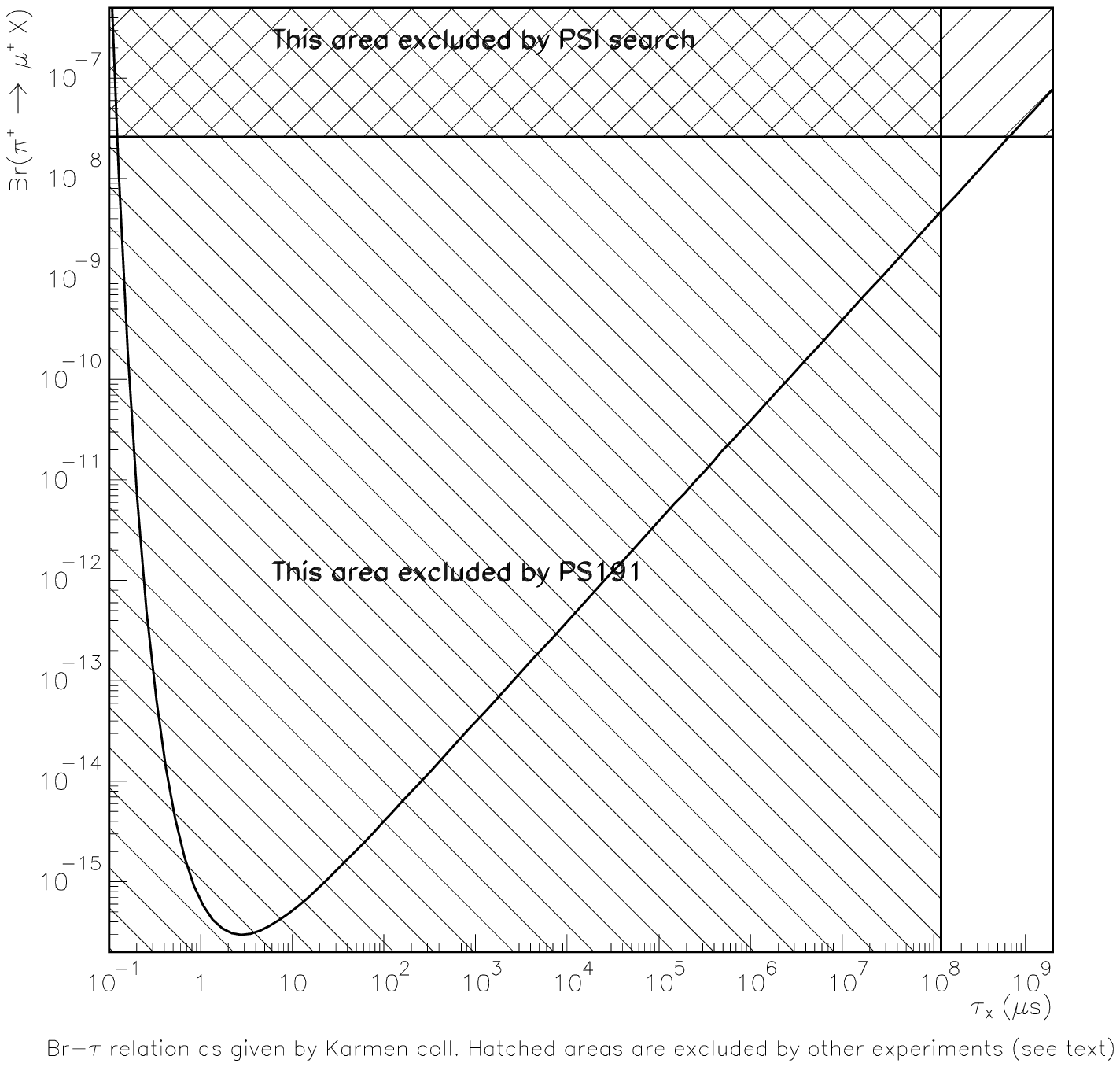}
\center{Fig. 1}
\newpage
\psfig{file=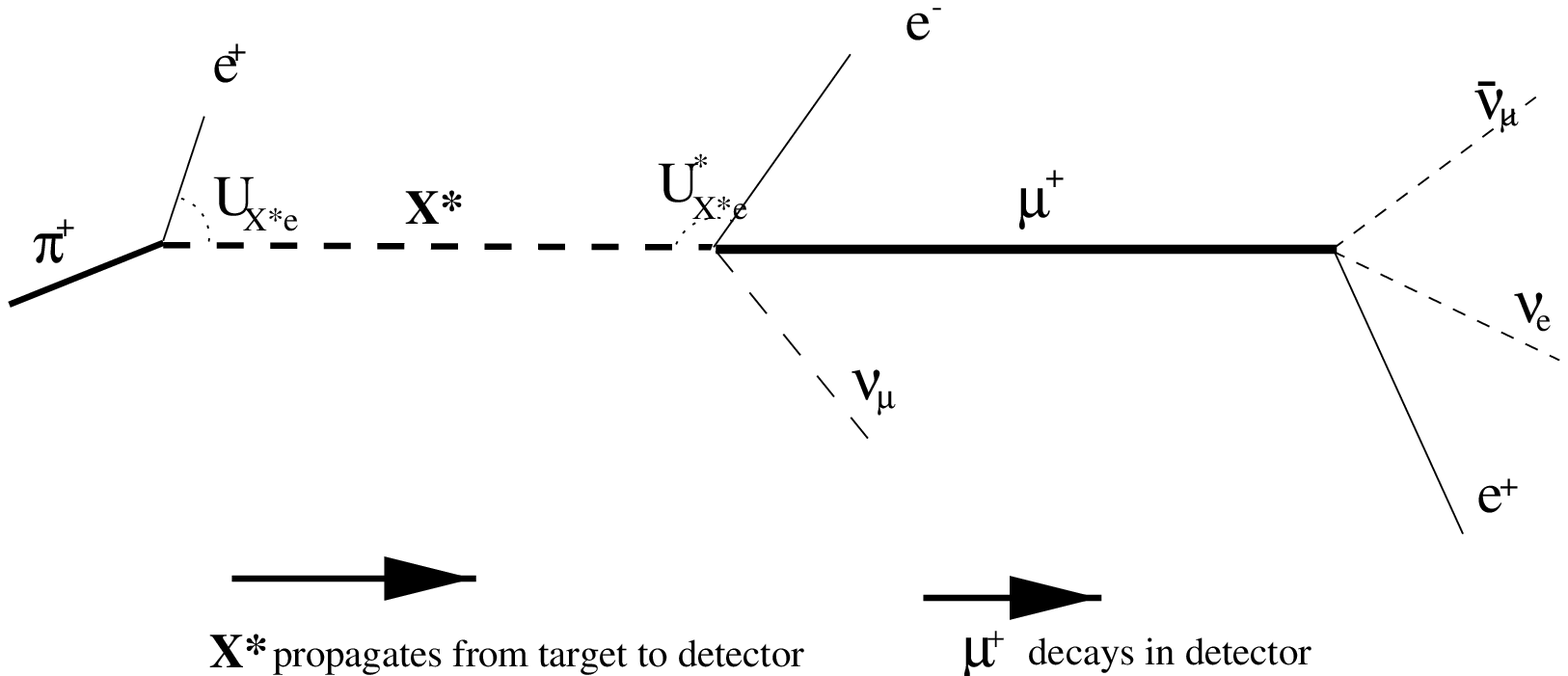}
\center{Fig. 2 The process hypothetized in this paper }
\newpage
\psfig{file=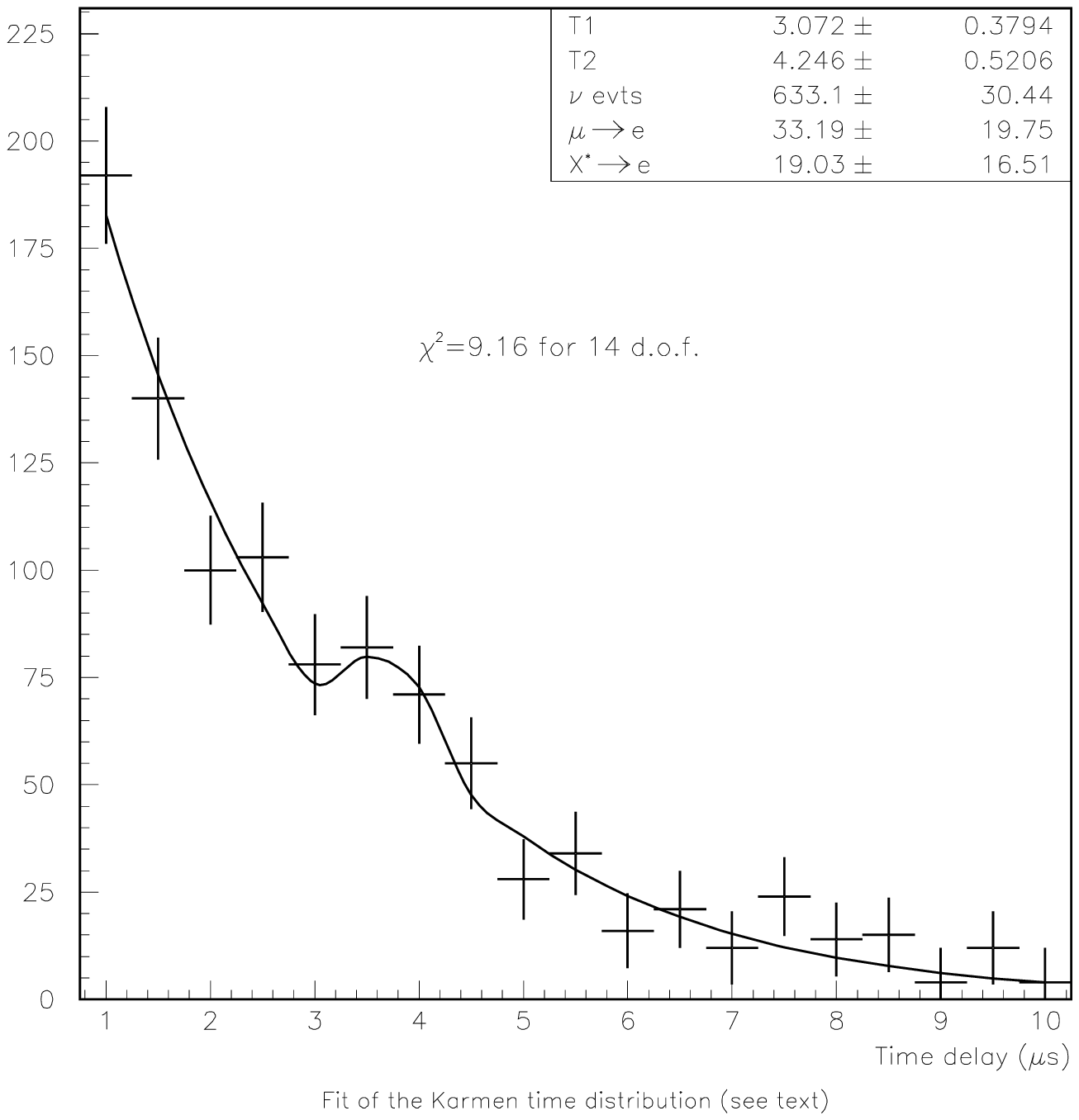}
\center{Fig. 3}
\newpage
\psfig{file=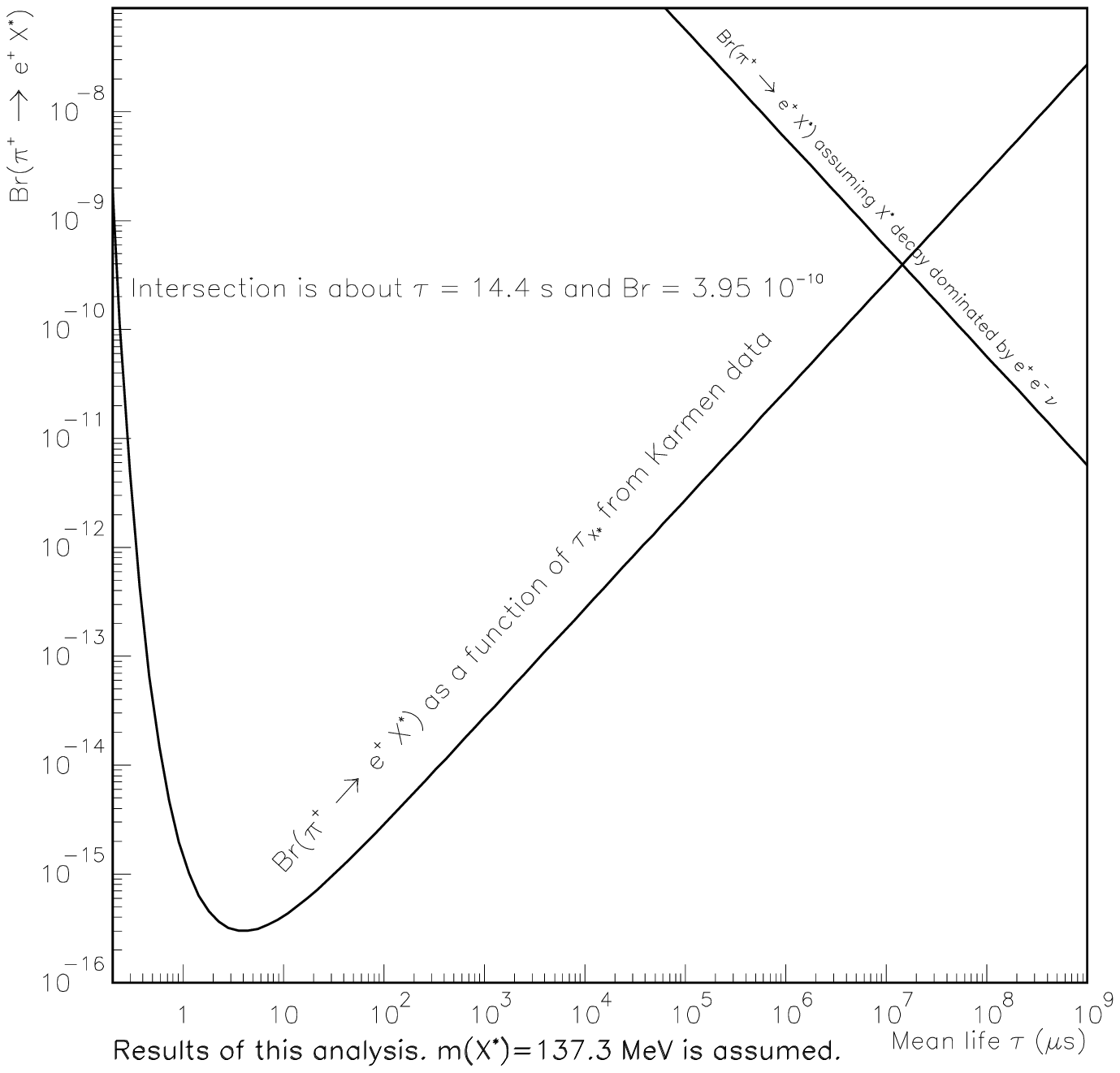}
\center{Fig. 4 }

\newpage
\end{document}